\documentclass[letter]{aa}
\usepackage{natbib}
\bibpunct{(}{)}{;}{a}{}{,}
\usepackage{graphicx}
\usepackage{txfonts}
\def\hcop  {\hbox{HCO$^+$}}      
\def\HCOP  {\hbox{HCO$^+$}}      
\def\Gmech {\Gamma_{\rm mech}}   
\begin{document}
   \title{Mechanical feedback in the molecular ISM of luminous IR galaxies}


\author{A.F.~Loenen\inst{1,2}, M.~Spaans\inst{1},  W.A.~Baan\inst{2} \and R.~Meijerink\inst{3}}
\authorrunning{Loenen et al.}
\offprints{A.F.~Loenen }

   \institute{Kapteyn Astronomical Institute, P.O. Box 800, 9700 AV  Groningen, the Netherlands \\
             \email{loenen@astro.rug.nl}
             \and
             ASTRON, P.O. Box 2, 7990 AA Dwingeloo, the Netherlands
         \and
         Astronomy Department, University of California, Berkeley, CA 94720
            }
   \date{Received 5 June 2008 / accepted 4 July 2008}

 \abstract{}{Molecular emission lines originating in the nuclei of
 luminous infra-red galaxies are used to determine the physical
 properties of the nuclear ISM in these systems. }{A large
 observational database of molecular emission lines is compared with
 model predictions that include heating by UV and X-ray radiation,
 mechanical heating, and the effects of cosmic rays.}{The observed
 line ratios and model predictions imply a separation of the observed
 systems into three groups: XDRs, UV-dominated high-density
 ($n$$\geq$$10^5$cm$^{-3}$) PDRs, and lower-density
 ($n$$=$$10^{4.5}$cm$^{-3}$) PDRs that are dominated by mechanical
 feedback.}{The division of the two types of PDRs follows naturally
 from the evolution of the star formation cycle of these sources,
 which evolves from deeply embedded young stars, resulting in
 high-density ($n$$\geq$$10^5$cm$^{-3}$) PDRs, to a stage where the
 gas density has decreased ($n$$=$$10^{4.5}$cm$^{-3}$) and mechanical
 feedback from supernova shocks dominates the heating budget.}


   \maketitle
%

\section{Introduction}
\label{sec:introduction}
Dense ($n$$>$$10^4{\rm cm}^{-3}$) molecular gas plays an important
role in the physics of (Ultra-) Luminous Infrared Galaxies. It
provides the materials from which stars are formed and fuels possible
active galactic nuclei.  The radiation originating in molecular gas
in the nuclei of these galaxies provides information about the
physical properties of the nuclear environment, such as the gas
density and chemical composition, and the dominant type of radiation
field. There are two important types of radiation: UV radiation
(6-13.6\,eV) in star-forming regions, which generates Photon Dominated
Regions \citep[PDRs; e.g.][]{1999RvMP...71..173H}; and X-rays
(1-100\,keV) emanating from accreting black holes, which produce in
X-ray Dominated Regions \citep[XDRs; e.g.][]{1996A&A...306L..21L,
1996ApJ...466..561M}.  Analysis of the state of molecular gas in a
significant number of galaxies allows systematic effects to be
identified, which can provide insight into the processes influencing
the (gas in the) nuclei of these galaxies: the star-formation rate,
the evolutionary stage of the system, and feedback processes.

\cite{2008A&A...477..747B} presented data for the CO, HCN, HNC, \HCOP,
CN, and CS line emission of 37 infrared luminous galaxies and 80
additional sources taken from the literature
\cite[e.g.,][]{NguyenEA1992, SolomonDR1992, AaltoPHC2002, GaoS2004a,
GraciaCarpioGPC2006}. They suggested that the sources can be divided
into PDRs and XDRs, based on the ratios of emission lines.  We expand
on their analysis and explore in more detail the different physical
and chemical processes responsible for the diversity in line ratios
found in the \cite{2008A&A...477..747B} data, by comparing
their data with the predictions of PDR and XDR chemistry models first
presented by \cite{MeijerinkS2005}.

\begin{figure*}[htbp]
\sidecaption
\includegraphics[width=12cm]{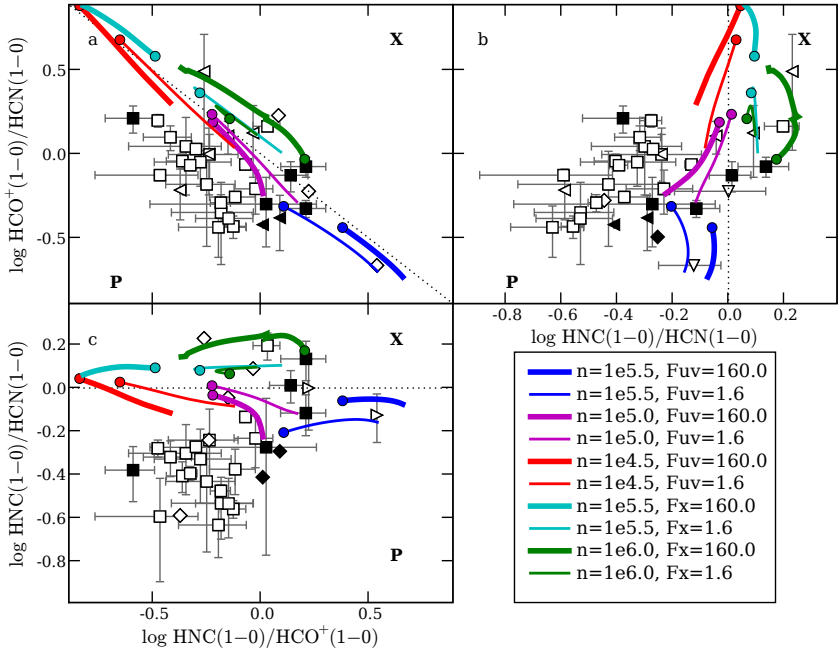}
\caption{Diagnostic diagrams using integrated line ratios of HCN, HNC,
and \HCOP\ versus each other. {\bf a)}~Integrated \HCOP/HCN versus
HNC/\HCOP\ ratios, {\bf b)}~Integrated \HCOP/HCN versus HNC/HCN
ratios, and {\bf c)}~Integrated HNC/HCN versus HNC/\HCOP\ ratios. The
symbols X and P and the dotted lines mark the regions of the line
ratios of XDRs and PDRs. Squares indicate reliable values and
triangles upper and lower limits.  In addition, filled symbols
indicate OH or H$_2$CO megamasers. Colored lines indicate results for
PDR and XDR modeling, as function of column density (see
sect.~\ref{sec:pdrs-xdrs}), where the dot denotes the highest
column. See Table~\ref{tab:shock-params} and legend for details about
the models; n denotes the density in cm$^{-3}$, and Fuv and Fx,
respectively, the UV and X-ray flux in erg s$^{-1}$ cm$^{-2}$.}
\label{3luik-no-shocks}
\end{figure*}

\section{PDR and XDR models}
\label{sec:pdrs-xdrs}

Figure~\ref{3luik-no-shocks} shows the HCN, HNC and \hcop\ line ratios
from \cite{2008A&A...477..747B}. To identify the physical processes
that generate the line ratios measured, we compare the data with the
predictions of a large number of PDR and XDR models.
\cite{MeijerinkS2005} and \cite{2007A&A...461..793M} presented a code
to model one-dimensional PDR and XDR clouds.  It takes into account
the various heating and cooling processes as well as the chemistry of
the gas. Because the excitation of the molecular species is included,
the intensities of individual molecular emission lines can be
calculated as well.  Our model predictions for both PDRs and XDRs are
shown in Fig.~\ref{3luik-no-shocks} by solid lines: the green and cyan
lines show the XDR results and the blue, purple, and red ones the
PDRs. These lines represent the line ratios emanating from different
depths in the cloud, and cover a column density range from about
N$_{\rm H}$$=$$10^{21.5}$ cm$^{-2}$ up to $10^{22}$$-$$10^{24}$ cm$^{-2}$
(depending on the density of the cloud).  Three conclusions can be
drawn from this comparison:

\noindent
{\bf 1:} PDRs and XDRs can be separated using the HNC/HCN line
ratio. In XDRs, HNC is always stronger than the HCN line, producing a
line ratio higher than unity, whereas for PDRs the inverse is
generally true (see Fig.~\ref{3luik-no-shocks}b,c).

\noindent
{\bf 2:} In the case of PDR chemistry, models with different densities
can be distinguished using the ratio of \hcop\ to HCN or HNC, which
decreases with increasing density (see Fig.~\ref{3luik-no-shocks}a).
This follows from the increased dissociative recombination rate of
\hcop with free electrons. A change in UV flux of two orders of
magnitude produces only modest changes in the line ratios, because the
UV field is significantly attenuated at the high column
densities where the molecules are abundant.

\noindent
{\bf 3:} The  most significant result from this comparison is that a
significant part of the data points has HNC/HCN and HNC/\hcop\ line
ratios that are systematically lower than those of the data points traced by
the models (log(HNC/HCN)$<$$-0.2$ and log(HNC/\hcop)$<$$-0.1$).  A lower
HNC/\hcop\ ratio can be obtained by decreasing the density of the gas
to $10^{4.5}$cm$^{-3}$ (red lines in Fig.~\ref{3luik-no-shocks}), but
the HNC/HCN ratios remain lower than can be explained by the PDR and
XDR models. \cite{MeijerinkSI2006b} investigated the effects of the
extra heating and ionization due to Cosmic Rays (CRs) and showed that
although CRs have an influence on \hcop, which is sensitive to the
ionization balance of the gas, they have little influence on the
abundances of HCN and HNC.  All this indicates that the state of the
gas is influenced by  processes that were not incorporated
into the \cite{2007A&A...461..793M} models.

\begin{figure*}[htbp]
\sidecaption
\includegraphics[width=12cm]{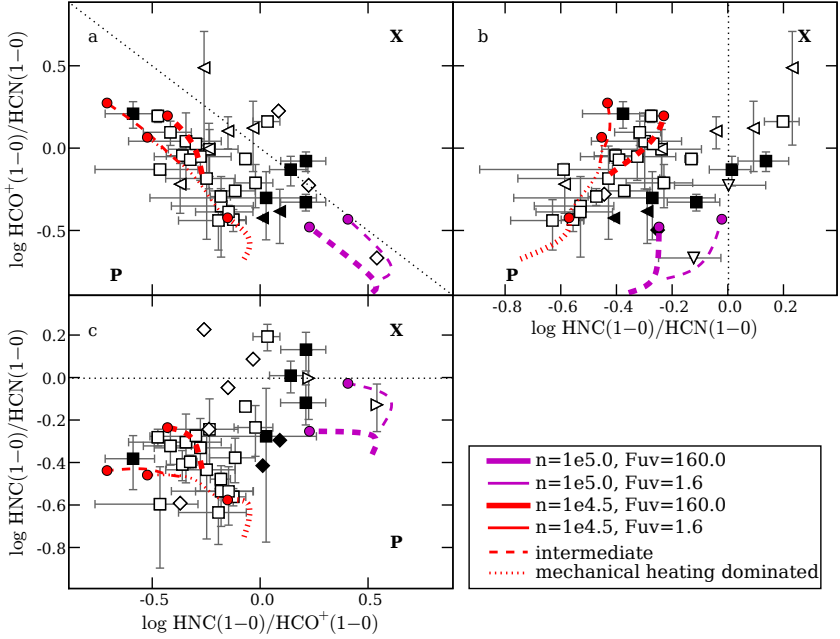}
\caption{ Diagnostic diagrams that show the effects of the added
mechanical heating.  See the legend and Table~\ref{tab:shock-params}
for details on the individual models.  Plot symbols are similar to
Fig.~\ref{3luik-no-shocks}. }
\label{3luik-shocks}
\end{figure*}

\begin{table}[t!]
    \caption{The important parameters for the
    PDR and XDR simulations and the line styles and widths used to plot the
    results in Figs.~\ref{3luik-no-shocks} and \ref{3luik-shocks}. }
    \label{tab:shock-params}
    \centering
    \begin{tabular}{lllllll}
      \hline
      \hline
      \noalign{\smallskip}
      &$n$        & \multicolumn{2}{l}{F} & \multicolumn{3}{l}{$\Gmech$ }\\
      & \multicolumn{1}{l}{[cm$^{-3}$]} & \multicolumn{2}{l}{[erg s$^{-1}$ cm$^{-2}$]} & \multicolumn{3}{l}{[erg s$^{-1}$ cm$^{-3}$]}\\
      \noalign{\smallskip}
      \hline
      \noalign{\smallskip}
       &         & \multicolumn{2}{l}{line width}             & \multicolumn{3}{l}{line style}\\
       &         & thin   &  thick                            & solid & dashed & dotted\\
      & & & & Fig.~\ref{3luik-no-shocks} & Fig.~\ref{3luik-shocks} & Fig.~\ref{3luik-shocks} \\
      \hline
      \noalign{\smallskip}
      PDR & $10^{4.5}$    & 1.6 & 160 & 0     & $3 \times 10^{-19}$ & $5 \times 10^{-19}$ \\
      & $10^{5.0}$    & 1.6 & 160 & 0     & $2 \times 10^{-18}$ & $9 \times 10^{-18}$ \\
      & $10^{5.5}$    & 1.6 & 160 & 0     & $9 \times 10^{-18}$ & $2 \times 10^{-17}$ \\
      \noalign{\smallskip}
      \hline
      \noalign{\smallskip}
      XDR& $10^{5.5}$    & 1.6 & 160 & 0     &- &- \\
      & $10^{6.0}$    & 1.6 & 160 & 0     &- &- \\

      \hline
  \end{tabular}
\end{table}

\section{Mechanical heating}
\begin{figure}[tbp]
\begin{center}
\includegraphics[width=\columnwidth]{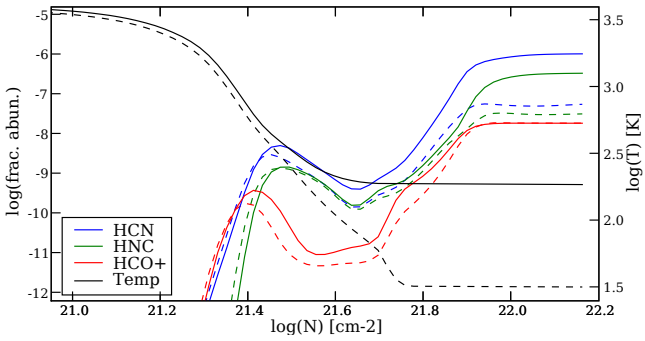}
\caption{Effects of mechanical heating on the temperature of the gas
and the abundances of HCN, HNC, and \hcop\ ($n$=$10^{4.5}{\rm
cm}^{-3}$, $F_{\rm UV}$=$160$ erg\,s$^{-1}$\,cm$^{-2}$,
$\Gmech$=$3\times10^{-19}$erg s$^{-1}$ cm$^{-3}$). Solid lines show
the results of a model with mechanical heating, dashed lines for a
model without. }
\label{abundances}
\end{center}
\end{figure}

The observed low HNC/HCN ratios could be produced by an enhanced
conversion of HNC into HCN in warm gas. At temperatures higher than
100\,K, HNC is converted efficiently into HCN
\citep[HNC+H$\rightarrow$HCN+H; ][]{SchilkeEA1992,
1996A&A...314..688T}.  Dense molecular gas temperatures in excess of
100\,K have been observed.  \cite{2005ApJ...629..767O} presented
interferometric maps of NH$_3$\ of the central 500\,pc of the nearby
starburst galaxy NGC\,253 and found clumps with kinetic temperatures
of between $\sim$150\,K and $\sim$240\,K. A similar study of H$_2$CO in
the starburst galaxy M\,82 by \cite{2007ApJ...671.1579M} found kinetic
temperatures of between $\sim$160\,K and $\sim$260\,K and densities on the
order of $10^4$ cm$^{-3}$.  We note that such temperatures will occur
only in the dense gas in the centers of the galaxies in our
sample. The large-scale low-density ($n$$<$$10^{4}$cm$^{-3}$)
molecular gas (as traced by CO) has lower temperatures.  In the PDR,
XDR, and CR models, temperatures of 100\,K or higher are only reached at
the edge of the clouds, where the abundances of HCN and HNC are too
low to be detectable. The regions deeper inside the clouds, where the
molecules become abundant (at N$_{\rm H}\geq 10^{21.5}$ cm$^{-2}$),
are cooler than 100\,K for radiative heating (see
Fig.~\ref{abundances}).

Since all radiative heating processes are insufficient to heat gas at
larger depths into the cloud, we add mechanical heating to the
models. Sources of this type of heating are discussed in the following
section. The mechanical heating is implemented by adding a depth
independent heating rate ($\Gmech$) to the total heating budget. A new
grid of models was calculated for several densities, radiation fields,
and mechanical heating rates (see Table \ref{tab:shock-params}) and
the results were plotted in Fig.~\ref{3luik-shocks}.  The adopted
mechanical heating rates were chosen in such a way that temperatures
deep inside the cloud (N$_{\rm H}$$\geq$$10^{21.5}$ cm$^{-2}$) were
between 150\,K and 250\,K. The mechanical heating rates are typically
about 10-50 times lower than the radiative heating rate at the edge of
the cloud, but start to dominate the heating balance deeper inside
the cloud (N$_{\rm H}$$\geq$$10^{21.5}$ cm$^{-2}$), where they are
several orders of magnitude higher than both the radiative and CR
heating rates.  Figure~\ref{abundances} illustrates how the
temperature increases when a mechanical heating term is added.  It
also demonstrates that the increase in temperature has the desired
effect on the abundances of the species: even though the total
abundances of both molecules in the cloud are enhanced, the abundance
of HCN increases with respect to HNC by a factor of about three.

The resulting line ratios are plotted in Fig.~\ref{3luik-shocks} and
show that, due to a decrease in the HNC/HCN ratio, data that could not
be explained previously are now described well by models with a
density of $10^{4.5}$cm$^{-3}$ and a range of UV fluxes and mechanical
heating rates (red lines).  Although models with higher densities
predict line strengths that are higher then the line strengths for the
$n$=$10^{4.5}$cm$^{-3}$ models and should be detectable, they produce
line ratios that are not observed (and therefore fall outside the
range of data points plotted in Fig.~\ref{3luik-shocks}).  An
explanation for why these high-density systems are not observed is
that the  mechanical energies needed to heat the gas are
unavailable (see Sect.~\ref{sec:sourc-mech-heat}).

\cite{2008arXiv0805.1801K} reported strong HCN emission in the cores
of some Seyfert galaxies (``HCN-enhanced nuclei'', or HENs).  Because
of the lack of evidence for a strong nuclear starburst (e.g.  lack of
mid-IR PAH emission), they attributed this to XDR chemistry.  However,
our results show that a weak PDR with mechanical heating can exhibit
the same characteristics.

\section{Sources of mechanical heating}
\label{sec:sourc-mech-heat}

The addition of mechanical heating to PDR clouds provides a
satisfactory means of explaining the observations. However, we need to
identify an available and sufficient source of energy. During the
evolution of a starburst, there are two feedback processes that
potentially provide sufficient mechanical energy to heat dense
molecular gas to the required temperatures. In the early stage of
nuclear star-formation (the first $10^5$$-$$10^6$ years), YSO outflows
inject mechanical energy into individual clouds. Then, after about
10\,Myr, the first massive ($M_\star$$>$$8M_\odot$) stars become supernovae
(SNe), and inject mechanical energy into the star-forming region,
which is distributed subsequently amongst the PDR clouds by turbulent
dissipation.  Since the energy injection required per cloud is known
from the chemical modeling ($\Gmech$), the required YSO outflow
velocities and SN rates for the different models can be calculated.

To estimate the effect of YSO outflows on the medium, we
follow the recipe of \cite{1990ApJS...74..833H} to determine the
energy injection rate needed to create an YSO outflow bubble as:
\begin{equation}
\frac{dE_{\rm flow}}{dt} \approx 3\times10^{29}r_{\rm pc}^2 v_{\rm flow}^3 n ~{\rm ergs\,s}^{-1}~,
\label{eq:1}
\end{equation}
where $r_{\rm pc}$ is the size of the bubble in pc, $v_{\rm flow}$ is
the outflow velocity in km\,s$^{-1}$, and $n$\ is the initial density
of the medium in cm$^{-3}$.  By assuming that the outflow energy is
distributed homogeneously over a cloud, of the same size as the
bubble, the heating rate is the energy injection rate divided by the
volume of the cloud: $\Gmech = \frac{dE_{\rm flow}}{dt} / V_{\rm
cloud}$.  From this, the outflow velocity needed to produce a certain
$\Gmech$ in a cloud can be calculated (using a typical cloud size of
0.1pc, which corresponds to a column density of N$_{\rm
H}$$\approx$$10^{22.2}$ cm$^{-2}$ and a extinction of
$A_V$$\approx$$15$ magnitudes) to be
\begin{equation}
v_{\rm flow} \approx \left( 4.1\times10^{25}\Gmech n^{-1} \right)^{\frac{1}{3}} ~{\rm km\,s}^{-1}~.
\label{eq:3}
\end{equation}
This yields  outflow velocities of around 10 km s$^{-1}$ (see
Table~\ref{tab:energy-budget}), which can be achieved in YSO
outflows. However, this  requires that every dense cloud has a YSO
nearby, which is unlikely since the YSO phase is short-lived.

To determine the SN rate required to heat clouds on a global scale
rather than for each individual cloud, we need to consider the size of
the total starburst region (a sphere of diameter $D_{\rm SB}$), and
the number of PDRs per kpc$^3$ ($f_{\rm PDR}$).  The transfer
efficiency ($\eta_{\rm trans}$) parameterizes the dissipation of
mechanical energy as heat as it cascades from the large scales of the
SNe shock bubbles to the individual clouds. The total energy injection
rate needed for the entire starburst region is then:
\begin{equation}
\frac{dE_{\rm SN}}{dt} = \frac{\pi}{6}  D_{\rm SB}^3 ~f_{\rm PDR} ~V_{\rm cloud} ~\Gmech ~{\rm ergs\,s}^{-1}~.
\label{eq:4}
\end{equation}
By assuming a mechanical energy output of $10^{51}$ ergs per SN, the SN rate
can then be expressed as:
\begin{equation}
{\rm SN~rate} = \frac{dE_{\rm SN}/dt}{3.17\times10^{43} ~\eta_{\rm trans}} ~{\rm SN\,y}^{-1}.
\label{eq:5}
\end{equation}
This yields SN rates ranging from $\sim$0.1 SN\,yr$^{-1}$ for the
$n$=$10^{4.5}{\rm ~cm}^{-3}$ models up to $\sim$6 SN\,yr$^{-1}$ for the
densest ($n$=$10^{5.5}{\rm ~cm}^{-3}$) clouds.  A more convenient measure
is to relate the SN rate with the star-formation rate (SFR), using the
initial mass function (IMF) $\Phi(M)$ \citep{1999A&A...350..349D}:
\begin{equation}
\frac{{\rm SN~rate}}{{\rm SFR}} = k = \frac{\int_{8M_\odot}^{50M_\odot} \Phi(M)dM}{\int_{0.1M_\odot}^{125M_\odot} M\Phi(M)dM}~.
\label{eq:6}
\end{equation}
%
\begin{table}[t!]
    \caption{ A summary of the results of the calculations of the
    energy requirements of the models, as described in
    Sect.~\ref{sec:sourc-mech-heat}.}
    \label{tab:energy-budget}
    \centering
    \begin{tabular}{lllll}
      \hline
      \hline
      \noalign{\smallskip}
      $n$       & $\Gmech$                 & $v_{\rm flow}$ & {\rm SN~rate}      & {\rm SFR} \\
      \multicolumn{1}{l}{[cm$^{-3}$]} & [erg\,s$^{-1}$\,cm$^{-3}$] & [km\,s$^{-1}$]  & [SN\,yr$^{-1}$] & [$M_\odot$\,yr $^{-1}$]\\
      \hline
      \noalign{\smallskip}
      $10^{4.5}$ & $3 \times 10^{-19}$ & 7.30  & 0.09 & 14.30  \\
                 & $5 \times 10^{-19}$ & 8.66  & 0.15 & 23.84  \\
      $10^{5.0}$ & $2 \times 10^{-18}$ & 9.36  & 0.61 & 95.35  \\
                 & $9 \times 10^{-18}$ & 15.46 & 2.75 & 429.07 \\
      $10^{5.5}$ & $9 \times 10^{-18}$ & 10.53 & 2.75 & 429.07 \\
                 & $2 \times 10^{-17}$ & 13.74 & 6.10 & 953.50 \\
      \hline
  \end{tabular}
\end{table}
Using a Salpeter IMF gives a value of $k=0.0064$. To estimate a value
for $f_{\rm PDR}$, we assume a region of size ($D_{\rm SB}$) 100 pc
and an ambient density of $10^{3}$$-$$10^{4}{\rm cm}^{-3}$, which
corresponds to a total mass for the dense molecular gas in the SB
region of $\sim 10^{6.5}$$-$$10^{7.5} M_\odot$. By assuming a clumping
factor of 10 and using the same cloud size and densities as before, we
obtain a $f_{\rm PDR}$of about 15 PDRs pc$^{-3}$. We note that only the
product of $D_{\rm SB}$ and $f_{\rm PDR}$ is important for the SN
rates and SFRs; a larger region with a lower PDR density will
yield the same results, since we determine the total molecular
gas mass that can be heated by SNe. The energy transfer efficiency
$\eta_{\rm trans}$ is assumed to be 10\%. A lower efficiency is not
expected, since the amount of energy coming from the star-formation
would otherwise become larger than the infra-red luminosities of the
sources.

Table~\ref{tab:energy-budget} lists the results of these
calculations. It shows that the models that trace most of the data
($n=10^{4.5}$cm$^{-3}$) require modest SFRs of only around 20
$M_\odot$ yr $^{-1}$, whereas the models with higher densities require
high SFRs.  Such high SFRs are difficult to achieve and will disrupt
the molecular region surrounding the star-forming region, which may
correlate with the fact that there are no sources observed with line
ratios corresponding to mechanically heated high-density PDRs.

Both YSO outflows and SNe can provide sufficient energy to heat the
gas, but SNe are more likely to be the dominant process for the
sources in our sample. First of all, SNe are able to inject the energy
on large scales, redistributing the energy  amongst the PDRs,
while YSO outflows are only able to heat single clouds. This requires
all stars and surrounding PDR clouds to be at the same (evolutionary)
stage.  Also, the time that an ensemble of stars exists in
the phase where SNe are produced is far longer than the YSO stage,
which lasts only $\sim 10^5$$-$$10^6$ years, implying that clouds
heated by SN feedback are far more likely to be observed.

\section{Conclusions}
\label{sec:conclusions}
Although PDR, XDR or CR chemistry models can explain the general
behavior of the HCN, HNC and \hcop\ line ratios  observed in
nearby star-forming galaxies \citep{2008A&A...477..747B}, they cannot
explain the low HNC/HCN line ratios observed in a large
number of systems. The addition of heating by mechanical feedback
provides a solution to this problem.

Our simulations in Figs.~\ref{3luik-no-shocks} and \ref{3luik-shocks}
show that the observed sources can be split into three main groups,
based on their HNC/HCN line ratios. First, there are a few sources that
exhibit XDR chemistry (log(HNC/HCN)$>$$0$).  The other (PDR) sources
can be divided into two groups: a small group (composed mostly of the
OH-MM and ULIRG systems) of sources that can be modeled with standard
PDR chemistry and high ($n$$\geq$$10^5$cm$^{-3}$) densities and a
larger group of sources with lower densities
($n$$\sim$$10^{4.5}$cm$^{-3}$) that are heated by mechanical feedback.

All evidence leads to the natural conclusion that the division between
the PDR sources is a result of the evolution of the star-formation
cycle of a galaxy, during which the dominant heating process changes
and density of the gas diminishes.  The high-density
($n$$\geq$$10^5$cm$^{-3}$) PDRs (with HNC/HCN ratios around unity and
weak \hcop\ lines) represent the early stage of star-formation where
the stars are forming in the densest clouds and the stellar UV
radiation dominates the chemistry in those clouds. During this period
most of the gas will be converted into stars and the increasing amount
of UV radiation of the stars will gradually penetrate a larger volume
of the (lower-density) gas.  The resulting lower average density leads
to an increase in \hcop\ with respect to HCN and HNC. After about
10\,Myr, the first massive stars become SNe and their shocks begin to
dominate the heating budget, leading to lower HNC/HCN line
ratios. These two effects alter the properties of the galaxies to
those of a second group consisting of mechanical-feedback-dominated
lower-density PDRs. Because of the short duty-cycle of the first stage
compared to the second, most sources in our sample are observed to be
in the later stage of their evolution.

It therefore appears that mechanical feedback plays a crucial role in the
time dependent physical state of the dense star-forming molecular gas
in luminous infrared galaxies.

\bibliographystyle{aa}
\bibliography{0327}

\end{document}